\documentclass[useAMS]{mn2e}
\usepackage{psfig}

\usepackage{times}

\def\bb{black--body}

\def\spose#1{\hbox to 0pt{#1\hss}}
\newcommand\lsim{\mathrel{\spose{\lower 3pt\hbox{$\mathchar"218$}}
     \raise 2.0pt\hbox{$\mathchar"13C$}}}
\newcommand\gsim{\mathrel{\spose{\lower 3pt\hbox{$\mathchar"218$}}
     \raise 2.0pt\hbox{$\mathchar"13E$}}}

\title[Supernova shock breakout in GRB 060218?]
{Did we observe the supernova shock breakout in GRB 060218? }

\author[Ghisellini, Ghirlanda \& Tavecchio]
{G. Ghisellini$^1$\thanks{E--mail: gabriele.ghisellini@brera.inaf.it},
G. Ghirlanda$^1$ and F. Tavecchio$^1$\\
$^{1}$Osservatorio Astronomico di Brera, via E.Bianchi 46, I-23807
Merate, Italy
}

\begin{document}


\pagerange{\pageref{firstpage}--\pageref{lastpage}} \pubyear{2002}

\maketitle

\label{firstpage}

\begin{abstract}
  The early optical data of GRB 060218 (first $10^5$ s after the trigger) have
  been interpreted as black--body emission associated with the shock breakout
  of the associated supernova.  If so, it is possible to infer lower limits to
  the bolometric luminosity and energetics of such a black--body component.
  These limits, which are rather independent of the emissivity time
  dependence, are tighter for the very early data and correspond to energetics
  $\sim 10^{51}$ erg, too large to be produced by the breakout of a supernova
  shock. A further problem of the above interpretation concerns the
  luminosity of the observed X--ray black--body component. It should be
  produced, in the shock breakout interpretation, as a \bb\ emission of
  approximately constant temperature from a surface area only slowly
  increasing with time.  Although it has been suggested that, assuming
  anisotropy, the long duration of the X--ray \bb\ component is consistent
  with a supernova shock breakout, the nearly constant size of the emitting
  surface requires some fine tuning.  These difficulties support an
  alternative interpretation, according to which the emission follows the late
  dissipation of the fireball bulk kinetic energy.  This in turn requires a
  small value of the bulk Lorentz factor.
\end{abstract}

\section{Introduction}

The observations of GRB 060218 by the Swift satellite (Gehrels et al.  2004)
have prompted Campana et al. (2006, hereafter C06) to interpret the presence
of a thermal, \bb, component in the soft X--ray band as the signature of the
shock breakout of the associated supernova (SN) 2006aj (e.g. Modjaz et al.
2006, Mazzali et al. 2006).  The low redshift ($z=0.033$, Mirabal et al.
2004), together with an unusually long prompt emission of GRB 060218, allowed
an unprecedented coverage by all of the three Swift instruments (UVOT, XRT and
BAT), providing simultaneous data from the optical to the soft $\gamma$--ray
range.  In the optical--UV band, the data in the different UVOT/Swift filters
showed a hard spectrum, which can be made consistent with a Rayleigh--Jeans
$F(\nu)\propto \nu^2$ law by invoking a Small Magellanic Cloud (SMC)
extinction law for the host absorption with $E(B-V)=0.2$ plus a galactic
$E(B-V)=0.14$ (C06).  In the 0.2--10 keV energy range the prompt spectrum can
be fitted by the sum of a $kT\sim 0.1$--0.2 keV \bb\ plus a cutoff power--law,
also consistent with the BAT 15--150 keV data.  A fit of the combined XRT and
BAT spectrum [0.2--150 keV], integrated over $\sim$ 3000 s, returns a peak
energy $E_{\rm peak}\sim 5$ keV, for which GRB 060218 is consistent with the
relation between the time integrated bolometric isotropic energy $E_{\rm iso}$
and the peak energy $E_{\rm peak}$ (Amati et al. 2002; Amati et al. 2007). GRB
060218 was underluminous, with $E_{\rm iso}$ slightly less than $10^{50}$ erg,
and in this respect it resembles GRB 980425 (associated with SN 1998bw; Galama
et al. 1998)\footnote{Note however that SN 2006aj is one order of magnitude
  weaker than SN 1998bw (Mazzali et al. 2006; Pian et al. 2006).}, GRB 031203
(associated with SN 2003lw; Malesani et al. 2004), and GRB 030329 (associated with
SN 2003dh; Stanek et al. 2003). From the temporal point of view, the time lag
between hard and soft emission is consistent (Liang et al.  2006) with the
lag--luminosity relation (Norris et al. 2000).

Two peaks are clearly observed in the optical--UV light curve of GRB 060218:
an initial flux increase lasting $3\times 10^4$ s is followed by a fast decay
until $t \sim 1.5\times 10^5$ s after the trigger; a second peak at $\sim$10
days shows the typical spectral signatures of the underlying SN (Ferrero et
al. 2006; Mirabal et al.  2006; Modjaz et al. 2006; Sollerman et al. 2006).
Spectroscopic observations indicated a time dependent expansion velocity:
2$\times 10^4$ km s$^{-1}$ at day 3, $\sim$1.8 $\times 10^4$ km s$^{-1}$ at
day 10 and a more rapid deceleration between day 10 and day 15 after explosion
(see Fig. 2 in Pian et al. 2006).  Polarization was detected a few days after
trigger (Gorosabel et al. 2006) at a level of a few per cent, indicating some
asymmetry of the emitting zone.
 
In the radio band, the flux between 2 and 22 days showed a typical
power--law decay ($\propto t^{-0.8}$, Soderberg et al. 2006).

The X--ray (0.3--10 keV) light curve presented a smooth long lasting
($\sim 3000$ s) peak followed by a fast decay. At 10$^4$ s the flux
began a shallower decreasing phase ($\propto t^{-1.2}$) lasting
several days. The spectrum of such phase is very soft, corresponding
to an energy spectral index $\alpha\sim 2.3$ (Cusumano et al. 2006).

\begin{figure}
\vskip -0.3 true cm
\centerline{ \psfig{file=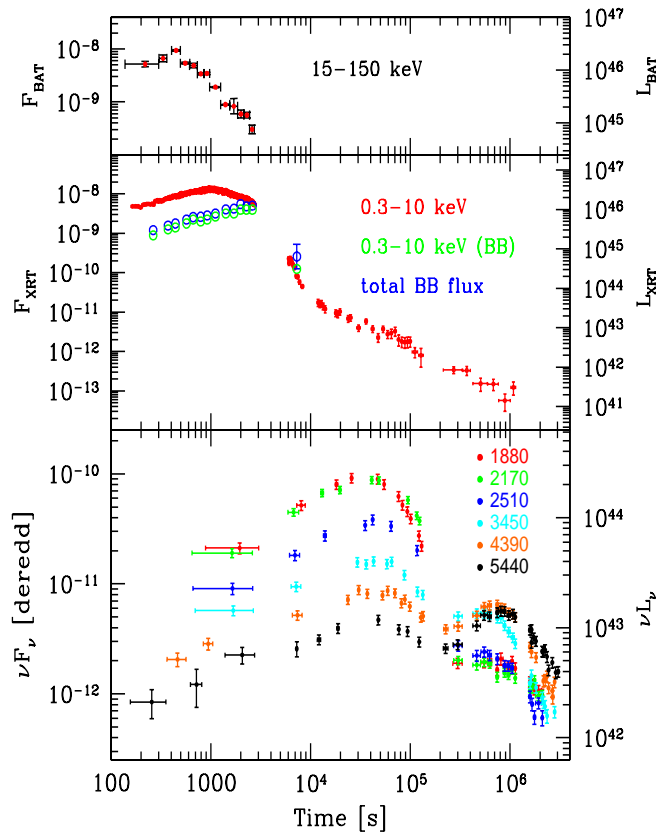,height=12cm,width=10cm} }
\vskip -0.5 true cm
\caption{Top panel: light curves of the total [0.3--10 keV] X--ray
flux of GRB 060218 (small red dots), the \bb\ flux in the same energy
range (green dots) and the bolometric flux of the \bb\ component (blue
dots). Data from C06 and W07. Bottom panel: the light curve in the
three UVOT/Swift optical--UV filters, de--reddened assuming a galactic
extinction $E(B-V)=0.14$ plus a host galaxy extinction $E(B-V)=0.2$
(for a Small Magellanic Cloud extinction curve for the latter).  cgs
units are adopted for all quantities.  }
\label{fig1}
\end{figure}

The complex behavior of GRB 060218 is summarized in Fig. \ref{fig1}, which
reports the available information on the light curves detected by all of the
three Swift instruments.  The top panel shows the light curve in the 15--150
keV band, as detected by BAT and analyzed by Toma et al. (2007). The middle
panel represents the 0.3--10 keV light curve as detected by XRT, and,
separately, the light curve corresponding to the \bb\ component only (as shown
by C06 and W07 too). The time dependent flux corresponding to the bolometric
\bb\ component is also plotted to show that its behavior reproduces that of
the 0.3--10 keV \bb\ light curve. Note that the \bb\ flux slightly increases
with time until $t\sim$ 3000 s, and that at 7000 s the absolute \bb\ flux has
decreased but its relative contribution to the total flux has increased. The
bottom panel reports the light curve in the optical--UV filters of UVOT. Note
that in C06 and in Waxman, Meszaros \& Campana (2007, W07 hereafter), no
absorption correction has been applied and the light curves refer to specific
fluxes multiplied by the FWHM of the different UVOT filters
[$F=F(\lambda)\Delta \lambda$]. Here we converted $F$ into the quantity $\nu
F_\nu$ ($\nu F_\nu=\lambda F(\lambda)$), and de--reddened the fluxes adopting, 
following C06,
$E(B-V)=0.14$ (galactic) plus $E(B-V)=0.2$ (host, with a SMC extinction law).


C06 interpreted the thermal X--ray spectral component, evolving towards
cooler temperatures and shifting into the optical/UV band, as emission
following the break out of a shock, driven by a mildly relativistic
shell, into the dense wind surrounding the progenitor.  Li (2007)
modeled numerically the corresponding transient emission specifically
for Type-Ibc SNae produced by the core collapse of WR stars
surrounded by dense winds. However, for the case of
GRB~060218/SN2006aj such a model required an unrealistically large
core radius of a WR progenitor star (but see W07).

The interpretation of the observational properties of GRB 060218
appeared therefore puzzling. Ghisellini et al. (2007, hereafter paper
I) discussed the alternative possibility that the optical--to--X--ray
radiation is non--thermal emission produced in a fireball moving with
a moderate bulk Lorentz factor ($\Gamma\sim 5$), and the thermal
X--ray component is due to some dissipation occurring within the jet
(possibly just below its photosphere).

Recently W07 fiercely argued against alternative interpretations by
presenting the details of the shock break out hypothesis in as
anisotropic SN explosion. According to W07, the anisotropy of the
explosion easily accounts for the long lasting X--ray thermal
emission.

In this letter we re-examine some aspect of the scenario proposed by W07. In
particular, following their interpretation, both for the optical and the
X--ray \bb\ components, we derive consequences which we see as its major
problems, as severe as to require alternative explanations. In Sec. 2 we point
out that, if the optical--UV emission belongs to a \bb\ component, the
evolution of the energy and temperature of the \bb, as inferred from the
available Swift/UVOT optical observations implies too large energetics for the
early phases.

It is then shown (Sec. 3) that 
the rather slow increase of the surface emitting 
the X--ray \bb\ is not a natural consequence of the
anisotropic shock breakout scenario, but rather favors an alternative
interpretation, where the emitting surface is associated with the
transparency radius of a relatively long lived, mildly relativistic
($\Gamma\sim$ a few) jet. Indeed, if the X--ray \bb\ is produced
following a funnel/jet shear instability (according to the ideas put
forward by Thompson 2006 and Thompson, Meszaros \& Rees 2007), then
the presence of the \bb\ component requires a small value of $\Gamma$,
of order $1/\theta_{\rm j}$, where $\theta_{\rm j}$ is the jet opening
angle.

\section{Black--body optical emission?}

We assume that the optical--UV emission up to $10^5$ s corresponds to
the Rayleigh--Jeans part of a \bb\ component, as suggested by C06 and
W07. For any assumed expansion law, it is then possible to estimate
the time dependence of the emitting surface.  Also the optical--UV
observations constrain the temperature dependence on time. Thus it is
possible to infer the bolometric luminosity and energetics. 

Consider the surface emitting as a \bb\ expanding with a velocity $v$,
starting from an initial radius $R_0$. In general, the expansion
velocity of the photosphere will decrease with time. The model adopted
by W07 postulates that the photospheric radius expands as $R \propto
t^{4/5}$, at least for $R>R_0$. Adopting the same dependence, the
Eq.~18 in W07 can be re-written as:
\begin{equation}
R \, =\, R_0 + 3.6\times 10^{10} t^{4/5} \quad {\rm cm}.
\label{r}
\end{equation}
This implies a photospheric expansion speed $v\propto t^{-1/5}$,
corresponding to the observed velocities 
(derived from spectral modelling)
2--3 days after the explosion (i.e. Eq. \ref{r}
gives $v=2.38\times 10^4$ km s$^{-1}$ after 3 days, in agreement with
the measurements reported by Pian et al. 2006). 

We then assume that for $t< 1.2\times 10^5$ s the optical--UV spectrum
corresponds to the Rayleigh--Jeans part of a \bb\ spectrum:
\begin{equation}
\nu F_\nu  \, = 2\pi^2 kT \left( {R\over d} \right)^2  {\nu^3\over c^2},  
\label{rj}
\end{equation}
where the distance $d=145$ Mpc.  This in turn defines the temporal
dependence of the \bb\ temperature:
\begin{equation}
T\, = \, {\nu F_\nu \over 2\pi^2 k}  \left( {d\over R} \right)^2  
{c^2 \over \nu^3}.  
\label{t}
\end{equation}
As the temperature decreases with time, the assumption that the
emission is in the Rayleigh--Jeans regime of the \bb\ spectrum is
appropriate, especially for the early data which set the most severe
constraints to the total energetics. Within the W07's scenario,
the \bb\ peak is entering into the UVOT band at $t\sim 10^5$ s, and
this implies that the temperature estimates at 1880 \AA\ is slightly
less than that estimated at all other wavelengths (see Fig.~2).

Fig. 2 summarizes our results, showing (from top to bottom), the time
profile of the assumed \bb\ radius (i.e. Eq. \ref{r}), temperature and
bolometric \bb\ luminosity. In the bottom panel it is also reported an
estimate of the \bb\ energetics, obtained by multiplying the
luminosity by the exposure time $\Delta t$ corresponding to each flux
measurement. The quantity $E_{\rm BB}= L_{\rm BB} \Delta t$ thus
provides a proxy of the implied \bb\ energetics.

The implied \bb\ luminosities are very large in the early phases,
being of the order of $10^{48}$--$10^{49}$ erg s$^{-1}$, implying \bb\
energetics $\sim 10^{51}$ erg (bottom panel). This is comparable to
the entire kinetic energy of SN2006aj, estimated to be $\sim 2\times
10^{51}$ erg by Mazzali et al. (2006).

We have considered different expansion laws, clearly bound by the
condition $v<c$, and other (reasonable) values for $R_0$. The result
does not change: at early times, i.e. for the smallest photospheric
radii, the observed flux requires high temperatures, and thus large
\bb\ luminosities.

{\it The resulting energetics are too large to originate as \bb\
emission from the heated SN envelope.} Therefore the observed flux is
not \bb\ radiation, though it might correspond to the absorbed
(Rayleigh--Jeans) portion of a spectrum becoming transparent below the
shortest observed wavelengths. Note that the $F(\nu)\propto \nu^2$
dependence somehow relies on the assumed host extinction: less
extinction implies a softer slope. Indeed, Sollerman et al. (2006)
argued for values $E(B-V)_{\rm Gal}=0.127$ and $E(B-V)_{\rm
host}=0.042$, significantly smaller than those adopted by C06 (who
required that the de--reddened optical--UV data follow a $F(\nu)
\propto \nu^2$ law). Assuming the $E(B-V)$ proposed by Sollerman et
al. (2006) we showed (in Paper I) that the optical--UV emission can
still be part of a synchrotron spectrum, partially self--absorbed,
connecting the optical-UV to the non--thermal X--ray
flux. Alternatively the optical--UV and X--ray fluxes could belong to
two unrelated components: in this case the optical--UV and X--ray
spectra should cut--off below 1880 \AA\ and $\sim 0.3$ keV,
respectively. Both interpretations minimize the required energetics.

\begin{figure}
\hskip -1 true cm
\psfig{file=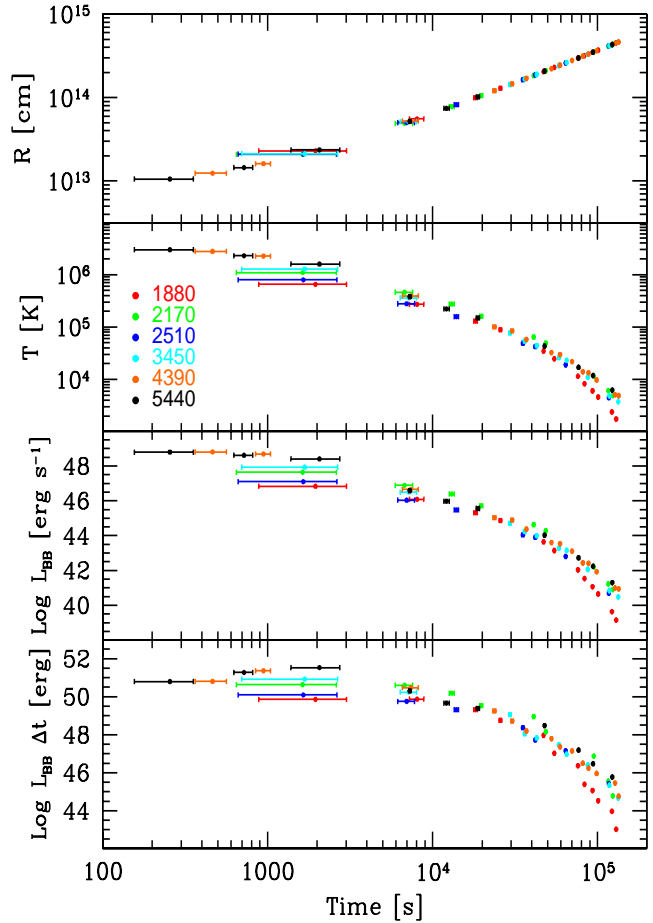,height=13cm,width=11.3cm} 
\vskip -0.5 true cm
\caption{From top to bottom: radius of the \bb\ emitting surface,
$R=R_0+3.6\times 10^{10} t^{4/5}$ with $R_0=7.5\times 10^{12}$ cm, as
suggested by W07; temperature of the \bb\ derived by assuming that the
whole spectrum belong to the Rayleigh--Jeans part of the \bb\
component; the corresponding \bb\ bolometric luminosity; \bb\
energetics, estimated as $E_{BB}=\Delta t L_{\rm BB}$, where $\Delta
t$ is the exposure time of each flux measurement. Different colors
refer to observations in different filters. }
\label{fig2}
\end{figure}

\section{The long lasting X--ray black--body}

The fluxes corresponding to the X--ray \bb\ component reported in C06
and W07 increase until $t\sim$ 3000 s. Such a component, albeit
fainter, is still detected at about $t\sim$ 7000 s. In Fig.~1 we
report the X--ray light curve as presented in C06 and W07 together
with the light curve inferred for the bolometric \bb\ flux to show
that they follow the same trend, namely $F_{BB}\propto t^{2/3}$.  The
radius $R$ of the (projected) emitting surface can be derived directly
from the fitting of the X--ray data, since the temperature of the
X--ray \bb\ is determined and the distance of GRB 060218 is known.
The temporal behavior of $R$ scales approximately as $R\sim a+bt \sim
5\times 10^{11}+3\times 10^8 t$ cm (see Fig. 3 in C06),
giving a rather modest expansion velocity, $v\sim 3\times 10^8$ cm
s$^{-1}\sim 0.01 c$. As the projected emitting surface scales as
\begin{equation} S_{\rm obs}(t) \, =\, 4\pi (a+bt)^2
\label{sobs} \end{equation} 
 the temporal dependence of the \bb\ temperature follows $T\propto
t^{1/6}/[5\times 10^{11}+3\times 10^8 t]^{1/2} \approx$ constant.

As mentioned, the puzzling properties of the X--ray \bb\ component, if
interpreted as produced by the SN shock breakout, are: i) the long
duration; ii) the large luminosity and energetics and iii) 
the slowly increasing emitting surface.
Li (2007) numerically examined the relevant
characteristics of shock breakouts in Wolf--Rayet stars with strong
winds, finding that although the presence of the wind allows to
radiate more efficiently, the resulting luminosities are still
insufficient to account for that observed from GRB~060218. Also the
predicted total duration of the event ($\sim$35 s) is much shorter
than observed ($\sim$ 3000--7000 s): indeed it is argued that possible
anisotropies -- although can partially increase the power generated --
cannot change the duration.

W07 instead claimed that the wind opacity, by leading to a
photospheric radius significantly exceeding the star radius, is the
key ingredient to explain the large luminosity. Furthermore, they
state that the long event duration is evidence for anisotropy, whose
origin however is different from that considered by Li (2007).
According to W07, because of the asymmetry the SN shock surface
reaches transparency at different times, depending on the direction,
thus accounting for the long lasting emission. Since the \bb\
luminosity increases with time, in this scenario the shock should
radiate more sideways than along the polar axis. Although this
interpretation is certainly possible, its realization requires some
fine tuning, as illustrated by the following simple example.

Assume for simplicity that the anisotropy concerns only the dependence
of the shock velocity on the angle $\theta$ from the polar axis and
that the whole emitting surface reaches transparency at the same
radius $R$ from the stellar center. The projected area at time $t$ is given 
by
\begin{equation} dS_\perp \, =\, 2\pi R^2
\sin\theta \cos\theta d\theta = -2\pi R^2 \mu {d\mu \over dt} dt.
\label{ds} \end{equation} 
The time derivative of Eq. \ref{sobs} leads to  
\begin{equation}
 8\pi b (a+bt) \, =\, -2\pi R^2 \mu {d\mu \over dt},
\label{ss}
\end{equation}
whose solution for $\theta$ is:
\begin{equation}
\sin^2\theta = {4 b^2 \over R^2 } t^2 +{ 8ab \over R^2}  t 
\to t={a \over b} 
\left[ \left( 1+{R^2 \over 4 a^2} \sin^2\theta \right)^{1/2} -1 \right]. 
\label{tsin}
\end{equation}
For $a=5\times 10^{11}$ cm and $b=3\times 10^8$ cm s$^{-1}$ the
condition $\sin\theta<1$ at 7000 s implies $R>5.1\times 10^{12}$
cm. 
Since the \bb\ emission lasts much longer than the characteristic 
light crossing time $R/c$, light travel time effects can be neglected.
Thus the elapsed time $t$ since the beginning of the \bb\ emission is 
\begin{equation} t \, =\, R \, \left( {1\over v(\theta)
} - {1\over v_0} \right) \, \to v(\theta) \, =\, {v_0 \over 1+ v_0
t/R}, 
\label{v} 
\end{equation} 
where $v_0\equiv v(\theta=0)$. Substituting Eq. \ref{tsin} in Eq. \ref{v} 
results in an almost constant velocity at small angles, decreasing
as $1/\sin^2\theta$ for larger $\theta$.

Different velocity profiles (or a dependence of the transparency radius on
$\theta$) would produce a different behavior of the measured emitting surface.  
Also, the observed increase of the \bb\ luminosity with time requires a
specific profile in $\theta$ for the energy released by the shock.

We propose that the nearly constancy of the \bb\ emitting
surface supports an alternative scenario, in which most of the emission 
is produced at the transparency radius of a ``GRB" jet. Models along these lines have
already been proposed by Thompson (2006) and Thompson, Meszaros \&
Rees (2007) in a different context: they suggest that the \bb\ 
component originates following shear instabilities between the jet and 
the funnel of the progenitor star and is of course released when the fireball 
becomes transparent. According to Daigne \& Mochkovitch (2002) this
occurs at a radius
\begin{equation}
R_{\rm ph}  \, \sim  4.6\times 10^{12}  {L_{\rm k, 48} \over (\Gamma/4)^3},  
\,\, {\rm cm}
\end{equation}
where $L_{\rm k}$ is the fireball kinetic power. 
The relevant radius for the 
\bb\ emission is then $\min [\theta_j, 1/\Gamma] R_{\rm ph}$, 
which is comparable to the radii inferred from observations if the jet is not highly 
collimated and/or the bulk Lorentz factor is small. The resulting isotropic \bb\ luminosity 
\begin{eqnarray}
L_{\rm BB, iso} & \sim & 4 \pi { R^2_{\rm ph} \over \Gamma^2} \sigma T^4
\nonumber \\ &\sim &
1.5\times 10^{46}  {L^2_{\rm k, 48} \over (\Gamma/4)^8}\, 
\left({ T \over 2\times 10^6}\right)^4 
{\rm erg~s^{-1}}
\label{lbb}
\end{eqnarray}
accounts for the observed one. 

Following these lines, in paper I we estimated the energy
requirements posed by assuming  that either the \bb\ represents the ``fossil"
radiation that accelerated the fireball in the first place, or it is 
produced at larger radii, following some dissipation event. 
The latter option was clearly favored: indeed in this case
the photon energies do not degrade significantly due to expansion between the dissipation and
the transparency radii, lowering the required energetics.

One possible origin of late dissipation is the process quoted above,
proposed by Thompson (2006) and Thompson, Meszaros \& Rees (2007),
namely shear instability between the fireball and the star funnel.
The resulting energy peak of the \bb\ spectrum was then associated with
the peak of the (time integrated) spectrum of the prompt emission, and
this allowed them to account for the Amati relation in terms of \bb\
emission. Note that GRB 060218 obeys the Amati relation, but only if
the peak energy of the overall X--ray spectrum is considered (i.e.
not the \bb\ peak energy)

Recently, by analyzing the sample of GRBs observed both by BATSE onboard
CGRO and by the Wide Field Camera onboard $Beppo$SAX, Ghirlanda et
al. (2007) found that the presence of a dominating \bb\ component faces
severe problems. Note that all of the GRBs considered follow the Amati
relation. Therefore the fireball/funnel instability, if occurs, may
not be responsible for the peak of the prompt spectrum, i.e. the bulk
of the emission. We clearly cannot exclude that 
it is responsible for \bb\ emission with lower temperature and luminosity, 
as observed in GRB~060218.

In all of the scenarios proposed to explain the peculiar
properties of GRB 060218, it is assumed a bulk Lorentz factor of order unity, i.e. 
a factor $\sim$100 smaller than the `canonical' value. The 
detection of the X--ray \bb\ might thus be associated with a small 
$\Gamma$ factor. Indeed, a key point in the scenario by Thompson et al. (2007)
concerns the value of $\Gamma$ required for efficient
dissipation: this has to be of the order of $\sim
1/\theta_j$. This may be the clue to understand why
\bb\ emission has been detected only in the spectrum of GRB 060218: 
for small--$\Gamma$ fireballs the shear instability may be efficient enough to
reveal itself through the presence of a \bb\ component, 
while for GRB fireballs with $\Gamma
\gg 1/\theta_j$ this should not be detectable. 

GRB 980425 is another event possibly characterized by a small $\Gamma$
fireball. Thus it is a likely candidate to show observable \bb\
emission in the soft X--rays\footnote{The possibility that the prompt
spectrum of GRB 980425 could have been similar to that of GRB 060218
has been discussed by Ghisellini et al. (2006).}, although it would
have been impossible to reveal it with the detectors in 1998.

\section{Summary and conclusions}

We have re--examined the possibility that the thermal components
detected in the early phases of the optical and X--ray emission
of GRB 060218 are due to the shock breakout of the associated SN 2006aj. 

We have found that if the optical--UV radiation corresponds to 
the Rayleigh--Jeans part of a \bb\ spectrum, the data imply very large \bb\
luminosities, especially at early times (i.e. the first few thousands
seconds after the trigger). The derived values cannot be accounted for
as emission by material in the envelope/wind of the star heated by 
the shock crossing.

Instead, the proposed interpretation that the optical--UV--to--X--ray
spectrum originates as non--thermal synchrotron emission appears
tenable and has two major advantages: (i) the optical--UV spectrum can
be softer than $\nu^2$  (allowing a smaller optical extinction,
as indicated by Sollerman et al. 2006) and (ii)
the energy requirement is significantly reduced with respect to the shock
breakout scenario.

The \bb\ spectrum detected in the X--ray band constrains the 
emitting surface to depend only weakly on time. 
The long total duration, which largely 
exceeds the light crossing time, led W07 to propose that the 
persistence of the \bb\ component 
is due to anisotropy, namely the fact that transparency is reached at 
different times by different parts of the shock surface, because
either the shock velocity or the photospheric radius
are functions of the polar angle.
Although this is a viable possibility, we have pointed out that 
only a specific dependence on the polar angle give rise 
to the observed behavior: in general, anisotropy implies that the 
resulting observed (projected) surface changes with time.

Instead, the almost constant X--ray emitting surface supports a
scenario in which the X--ray \bb\ emission is produced inside a jet, 
before transparency is reached. The transparency radius corresponds to 
that inferred from observations if the bulk Lorentz factor is small, of 
the order of a few. This is also the condition to develop efficient shear
instability modes between the jet and the funnel, and explains why the
X--ray \bb\ component can be rarely observable (i.e. only when the bulk
Lorentz factor is of the order of the inverse of the jet opening
angle). 

We noted that GRB 060218 obeys the $E_{\rm peak}$--$E_{\rm iso}$ (Amati) relation 
only if the entire (non--thermal?) X and $\gamma$--ray emission component, 
which peaks at $\sim 5$ keV, is considered (the X--ray thermal emission peaks
at too low energies). If this is not a coincidence, it implies that the physical process
underlying the Amati relation is robust and independent of 
the bulk Lorentz factor. Similar considerations hold for the
lag--luminosity relation, which GRB 060218 obeys as well.

The bottom line of this re--analysis is that we are still puzzled
about the self--consistency of the SN shock breakout interpretation 
of the optical, UV and thermal X--ray emission of GRB~060218.

On the other hand, this bursts $is$ associated with SN 2006aj, and
therefore some signs of the associated shock breakout should be present. Why
should we not observe it? The simplest answer is that the emission associated 
with the shock break out is weaker than other spectral components.


\section*{Acknowledgements}
This work has been partially funded by a 2005 INAF grant.  We thank
the anonymous referee for useful criticism. We are grateful to
Annalisa Celotti for comments and discussion.

\end{document}